\documentstyle[aps,prl,floats,epsfig]{revtex}

\title{Symmetric Versus Nonsymmetric Structure of the Phosphorus Vacancy on InP(110)}
\author{Ph. Ebert$^1$, K. Urban$^1$, L. Aballe$^2$, C. H. Chen$^2$, K. Horn$^2$, G. Schwarz$^2$, J. Neugebauer$^2$, and M. Scheffler$^2$}
\address{
$^1$Institut f\"ur Festk\"orperforschung, Forschungszentrum J\"ulich GmbH, D-52425
J\"ulich, Germany
}
\address{
$^2$Fritz-Haber-Institut der Max-Planck-Gesellschaft, Faradayweg 4--6, D-14195 Berlin-Dahlem, Germany
}

\begin{document}
\draft
\twocolumn[\hsize\textwidth\columnwidth\hsize\csname@twocolumnfalse\endcsname

\maketitle

\begin{abstract}
The atomic and electronic structure of positively charged P
vacancies on InP(110) surfaces is determined by combining scanning
tunneling microscopy, photoelectron spectroscopy, and density-functional
theory calculations. The vacancy exhibits a nonsymmetric rebonded
atomic configuration with a charge transfer level $0.75 \pm 0.1$ eV above the
valence band maximum. 
The scanning tunneling microscopy (STM) images show only a time average of two
degenerate geometries, due to a thermal flip motion between the mirror
configurations.  
This leads to an apparently symmetric STM image, 
although the ground state atomic structure is nonsymmetric.
{\em Copyright 2000 by the American Physical Society.}
\end{abstract}

\vskip2pc]

Although it is well known that point defects can exert a profound
influence on the physical properties of semiconductors, the
determination of the atomic scale geometric and electronic structure of
point
defects has remained a challenging task for theoretical as well as experimental
research. One particularly striking example is anion
vacancies on (110) surfaces of III-V semiconductors, where no
agreement has been reached regarding even basic properties, such as (i) the
symmetry of the atomic structure and (ii) the energy of defect levels in the
band
gap: Although recent density-functional theory (DFT) calculations for
positively
charged arsenic vacancies on the GaAs (110) surface agreed that the gallium
atoms neighboring the vacancy relax into the surface layer (in contrast with
an
earlier tight-binding calculation \cite{4}), one calculation found a rebonded
configuration breaking the mirror symmetry of the surface \cite{1}, whereas
the other predicted a fully symmetric configuration to have the lowest energy
\cite{2}.
High resolution scanning tunneling microscopy (STM) images show a density of
states preserving the mirror symmetry of the surface at the defect site
\cite{4,7}.
Although this seems to favor a symmetric atomic structure, the experimental
results could not be matched to results of any of the DFT
calculations\cite{9,10,11,12}. 
Furthermore, scanning tunneling spectroscopy (STS) yielded a local downward band
bending of 0.1 eV \cite{4}, whereas surface photovoltage measurements found a
band bending of $0.53\pm 0.3$ eV \cite{19} at the site of the positively charged
As
vacancy on $p$-doped GaAs(110) surfaces. 
Concerning the energy levels, the two
different DFT calculations predicted the charge transfer levels (+/0) to be 0.32 eV
\cite{1} and 0.1 eV \cite{2}, and the lowest Kohn-Sham
eigenvalues in the band gap to be  0.73 eV \cite{1} and 0.06 eV \cite{2} above the
valence band maximum (VBM). 
In view of this puzzling situation it is obvious that the interpretation of the
experimental STM images as well as the structure of the anion vacancies on
(110) surfaces of III-V semiconductors are still under debate
\cite{9,10,11,12}.

    Discrepancies such as those pointed out above can arise due to limitations
of the methods used. On the theoretical side DFT calculations
have proven to be powerful to determine the structure of point defects \cite{20}. 
Yet differences in the size of the supercell, the
pseudopotentials employed, or the exchange-correlation functional implemented 
might lead to different results, such as the defect's symmetry or the position 
of the defect levels, especially in situations where two configurations are 
almost degenerate in energy. 
On the experimental side the STM images alone do not provide
direct information about the atomic structure nor are the experimental
conditions, notably the concentration of the vacancies, the same in all
experiments.
Furthermore the quantities discussed to describe the position of the defect level,
i.e., Kohn-Sham eigenvalues within DFT, band bending, and charge
transfer levels, are 
different physical quantities and thus (typically) deviate from each other
significantly. 
In addition, the tip of the STM may affect the structure of the vacancy, since
it is well known that the tip can even excite vacancies to migrate \cite{8}. 
Therefore the combination of state-of-the-art theory and \emph{well defined}
experiments is needed to get reliable results and test the correct
interpretation of each of the separate results. 

In this letter we combine three different methods to determine the geometric
and electronic structure of phosphorus (P) vacancies on $p$-doped InP(110) 
surfaces:
STM, photoelectron spectroscopy (PES), and DFT. This enables us
to determine the charge transfer level of the defect state in the band gap and
the equilibrium atomic structure of the vacancy in $p$-type material. We
identify the rebonded structure as the ground state of the positively charged
defect and demonstrate that a thermal flip motion between two mirror
configurations results in simulated STM images consistent with the
experimental findings.  
Our results solve the apparent discrepancies of the previous experimental 
and theoretical studies~\cite{110}, provide a comprehensive picture of the vacancy
structure, and point out a methodology to determine the structure of defects
with high accuracy.

%FIG. 1 \\ 
\begin{figure}[t]
\center{
\epsfig{figure=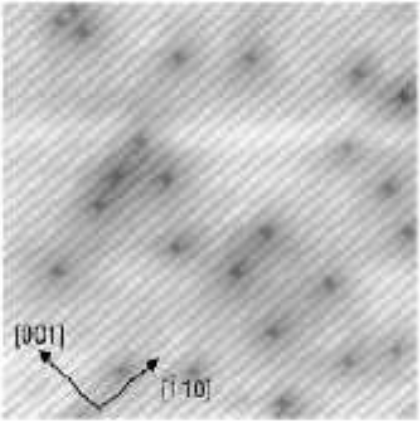,width=56.47mm}
}
\caption[]{
Occupied state STM image of P monovacancies formed by
annealing $p$-doped InP(110) surfaces at $\rm 160^{\circ}C$. The image has
been acquired at $-2.7$ V tunneling voltage and 0.9 nA current.
}
\end{figure}

The surface vacancies were produced in high concentrations by annealing 
initially defect-free $p$-doped InP(110) cleavage surfaces [carrier
concentration $n_{\rm dop}$ of $(1.3-2.1)\times 10^{18} {\rm cm}^{-3}$] 
at  $\rm 160^{\circ}C$ \cite{13,14}. 
Figure 1 shows that this procedure leads indeed to only one
type of defect on the surface which was identified as a singly positively
charged phosphorus vacancy previously  and whose concentration can be
controlled by the  annealing time \cite{13,14,18}. 
We prepared several samples from the same wafer
with a variety of annealing times in an ultrahigh vacuum (UHV) chamber
at the "Berliner Elektronenspeicherring f\"ur Synchrotronstrahlung"
storage ring for the photoelectron spectroscopy and in another UHV
system for the STM measurements ensuring an exact control of defect
concentration in both experiments. 
All measurements were done at room temperature.
Angle-resolved PES measurements were used to probe the valence band in 
normal emission at a photon energy of 50 eV, and the indium $4d$ core level.
The InP sample was in Ohmic contact with a Au metal surface cleaned by
scratching, on which the position of the Fermi level was measured. With
the STM we measured constant-current images at a variety of voltages.
From these images the vacancy concentrations were determined as a
function of the annealing time \cite{13} as well as the spatial distribution
of the occupied and empty density of states of the vacancies with
atomic resolution. 

We first focus on the determination of the {\em energy of the defect state}.
Here we concentrate on the {\em charge transfer level} of the surface vacancy,
which is defined as the position of the surface Fermi level at which the defect 
changes its charge state.  
This energy level can be calculated by DFT and can be probed by measuring the
band bending  as shown below.  
We note that the energy of the charge transfer level might be very different
from that of the ionization levels or the Kohn-Sham eigenvalues within DFT. 
In particular, large deviations between these quantities might occur in
systems with large atomic relaxations such as surface and bulk defects \cite{20}. 

STM images reveal that during heat treatment of initially
defect-free cleavage surfaces the vacancy concentration increases with
increasing annealing time.  
Similarly, the position of the Fermi level, which on well-cleaved defect-free
surfaces is close to the top of the valence band (VBM) for $p$-type samples,
is found to shift toward midgap upon annealing, as measured by PES (see Fig. 2). 
The filled diamonds (left axis) show the band bending at the surface, 
as determined through PES, whereas the open circles show the vacancy
concentration (from STM, right axis) as a function of the annealing time. 
The band bending reaches a saturation value of 0.65 eV at vacancy
concentrations of $5\times 10^{12} {\rm cm}^{-2}$. 
Figure 2 also demonstrates that the Fermi level shift is directly correlated
to the increase of the vacancy concentration. 

By combining the vacancy concentration and the band bending it
induces, it is possible to determine the energy of the charge transfer level in
the band gap $E_{sd}$, because the charge per surface area in the surface
layer $Q_{ss}$ is exactly compensated by the charge density per surface area
in the depletion layer $Q_{sc}$ \cite{15}. 
The charge per surface area induced by a concentration $n_{sd}$ of $+1e$ charged
P surface vacancies is \cite{15} 

%EQUATION 1
\begin{equation}
 Q_{ss} = \frac{e\, n_{sd}}{{\rm exp}((E_F - E_{sd})/kT)+1} \ ,
\end{equation}

\noindent
where $E_F$ is the Fermi energy. 
The difference in energy of the charge transfer level and the Fermi level  
$E_{sd}-E_{F}$ is given by
$(E_{sd}-E_{sv})-(E_v-E_{sv})-(E_F-E_v)$ (for definitions, see inset in Fig. 2; 
$E_v$ and $E_{sv}$
are the positions of the valence band in the bulk and at the surface,
respectively) where $E_v-E_{sv}$ is the band bending $eV_s$ measured by
photoelectron spectroscopy at the surface. The charge density $Q_{sc}$ in
the depletion layer compensating  $Q_{ss}$ is

%EQUATION 2
\begin{equation}
 Q_{sc} = \sqrt{2 \epsilon_0 \epsilon_r n_{\rm dop} kT
                \left [ {\rm exp}(\frac{-eV_s}{kT})+\frac{eV_s}{kT}-1) \right ]}
 \ .
\end{equation}

The energy difference between the Fermi energy and the valence band
maximum in the bulk has been determined for a nondegenerate InP
crystal from the carrier concentration according to 

%EQUATION 3
\begin{equation}
 E_F - E_v = -kT\  {\rm ln} \left [ 4 n_{\rm dop}\, (\frac{2m_p\, kT}{\pi
 \hbar^2})^{-3/2} \right ]
\end{equation}

\noindent
with $m_p$ being the mass of the holes.

These equations permit a determination of the difference in energy
of the charge transfer level $E_{sd}$ and the valence band maximum at the
surface $E_{sv}$ using the vacancy concentration dependence of the band
bending (see Fig. 2). We obtain an energy for the charge transfer level of the P
vacancy of $0.75 \pm 0.1$~eV above the valence band maximum at room
temperature.
Note that the knowledge of the defect concentrations is crucial
to determine the charge transfer level from a measurement of the band
bending.

%FIG. 2 \\
\begin{figure}%[b]
\center{
\epsfig{file=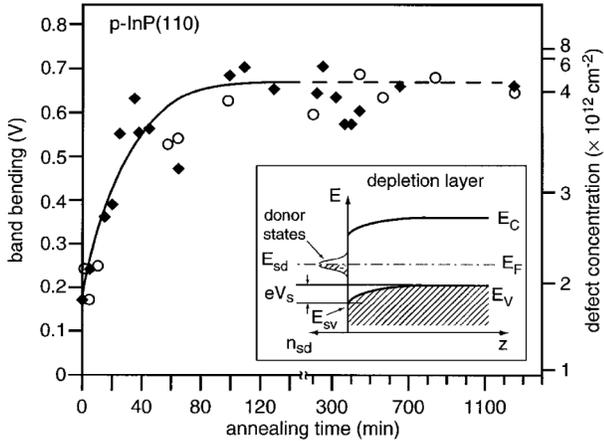}
}
\caption[]{
The band bending (filled diamonds, left axis) and the vacancy
concentration (empty circles \cite{13}, right axis) on the InP(110) surface
as a function of the annealing time at $\rm 160^{\circ}C$. The solid line is the
vacancy concentration calculated on the basis of Eqs. (1) and (2) and a fit
through the measured band bending values. The dashed line shows the
saturation value. The right scale is chosen such that according to Eqs.
(1) and (2) the vacancy concentration is shown on a scale linear
proportional to the band bending. The inset shows the schematic drawing
of the depletion region in case of a fully pinned surface.
}
\end{figure}

At this stage we compare this energy value with results from our
total-energy density-functional theory calculations. 
Details of the calculational method used are described in Refs. \cite{3,17}.
Here we only increased the cutoff of the plane wave basis set to 15 Ry in
the electronic structure calculation to ensure convergence of the indium
pseudopotential. We used a $2 \times 4$ and 6 layers thick supercell,
which is large enough here, since the calculated bond lengths changed less
than 0.01 \AA\ and the defect formation energy  by less than 0.01 eV
when increasing the size of the supercell.
For the $+1e$ charged P vacancy the calculation yields two different
equilibrium atomic configurations, whose total energies differ by only 0.07
eV. 
The vacancy configuration with the lowest energy exhibits a rebonding
of one of the neighboring surface atoms with the indium in the second
layer leading to a nonsymmetric atomic structure with respect to the
($1\overline{1}0$) mirror plane of the defect-free (110) surface [see atomic
structure superposed on Fig. 3(a)]. 
The calculated indium-indium spacing of this configuration is 2.89 \AA\ as
compared to 2.98 \AA\ for the symmetric ground state structure of the neutral
defect. A similar relaxation has also been calculated for the positively
charged As vacancy on GaAs(110) \cite{1,110}.
From our calculations we find a charge transfer level (+/0) at 0.52 eV above
the VBM for the nonsymmetric vacancy. 
For the energetically less stable symmetric configuration this level is at 0.45 eV. 
Both values deviate from the measured charge transfer level by roughly 0.2 eV.
We attribute this deviation to the well-known underestimation of energy levels
in local density approximation. 
An approximate upper limit of the error is the difference between the
calculated and experimental band gap. 
Using our calculated value (1.26 eV) and the experimental one (1.42 eV) we
get a systematic error of about 0.16 eV.
We can therefore conclude that the theoretical and experimental positions of
the defect levels agree well within the error margins.  
We note, however,  that the systematic error for the calculated energies of
the charge transfer levels is too large to identify the symmetry of the  vacancy
on the
position of the defect level only. 
In the following we will show that an unambiguous identification can be
obtained based on a comparison of the total energies and an analysis of the
STM micrographs. 

%FIG. 3 \\
\begin{figure}[t]
\center{
\epsfig{width=81.2mm,file=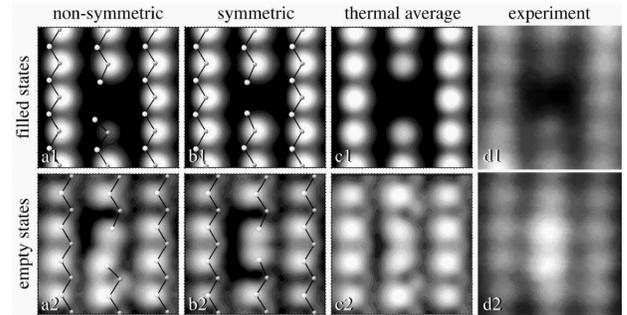}
}
\caption[]{
Simulated STM images deduced from the calculated local
density of states 3 \AA\ above the surface of a $+1e$ charged P vacancy
on InP(110) surfaces for different atomic configurations (a)$-$(c) and
experimental high resolution STM images of a P vacancy measured at (d1)
$-2.9$ and (d2) $+2.0$ V. The simulation shows the occupied (upper frames,
energy range from VBM to VBM $-1$ eV) and empty (lower frames, energy
range from VBM to VBM $+2$ eV) states for (a1),(a2) a nonsymmetric
rebonded; (b1),(b2) a symmetric nonrebonded; and (c1),(c2) the time
average of the two mirror configuration of the nonsymmetric rebonded
vacancy configuration. The respective atomic configurations are
superposed on the simulated images.
}
\end{figure}
 
We now focus on the {\em atomic structure}. STM images of $+1e$ charged P
vacancies [Fig. 3(d)] always exhibit an apparently symmetric density of states
with respect to the ($1\overline{1}0$)  mirror plane. 
The STM images thus suggest a  symmetric vacancy structure
although in our calculations we find the nonsymmetric structure to be lower 
in energy.
In order to clarify this apparent conflict, we have calculated in Fig.
3(a)$-$(c) simulations of occupied (upper frames) and empty (lower
frames) state STM images from the local density of states \cite{3} of 
different vacancy configurations. 
Frames 3(a1) and 3(a2) show the simulated STM images for the nonsymmetric case,
whereas frames 3(b1) and 3(b2) show the same for the symmetric vacancy
configuration.  
Both vacancy configurations lead to clearly distinguishable STM images: 
the nonsymmetric vacancy exhibits a pronounced asymmetry. 
The P atom next to the rebonded indium atom [upper atom in Fig. 3(a)] 
has a brighter dangling bond than the P atom at the opposite side of the vacancy
and the rebonded indium atom has a weaker maximum in the empty states. 
Lateral displacements are also visible. 
These effects are purely electronic, because the calculated height 
difference is only 0.11 and 0.07 \AA\ for the cations and anions, respectively. 
In contrast, the symmetric vacancy shows the expected symmetry with respect to
the ($1\overline{1}0$) mirror plane [Fig. 3(b)]. 
The images of the occupied states exhibit very little changes of the
neighboring occupied dangling bonds [Fig. 3(b1)], and the separation of the
two neighboring empty dangling bonds is considerably reduced [Fig. 3(b2)].
If we compare these simulations with the experimental STM images we find a
symmetric density of states [Fig. 3(d)] in conflict with the energetically
favorable nonsymmetric vacancy configuration in Fig. 3(a). 
However, also the symmetric vacancy configuration does not agree with the 
experimental STM images:
The empty state STM images do not show any reduced separation of the two
neighboring empty dangling bonds in contrast to the simulation of the
symmetric vacancy structure [Fig. 3(b2)].
Thus none of the static equilibrium vacancy configurations can be matched in
detail to the experimental observation. 

This conflict can be resolved by closer inspection of the  barrier between 
the two possible mirror configurations of the asymmetric vacancy.
We estimated an upper limit for this barrier by mapping the energy along the
direct reaction path between the symmetric and nonsymmetric vacancy
configurations. 
On this basis we found an upper limit of $0.08$ eV for the barrier between the
two nonsymmetric configurations.
The low value of the barrier implies that, at room temperature, the vacancy 
flips between the two nonsymmetric configurations at a rate significantly 
higher than the time resolution of the STM (1 to 10 Hz). We estimate a
flip rate at
room temperature of 0.1 THz using the transition state theory, the optical
phonon frequency of 2 THz \cite{100} as attempt frequency, and the calculated 
barrier. 
This concept of a thermally activated flip motion has been well established 
for dimers at the Si(001) surface \cite{16}. 
Based on these findings we interpret the STM image as a time average of the 
flipping vacancy.

In order to test the model of a thermal flip motion we have
performed STM image simulations based on this flip mechanism. In Fig. 3(c) we
assumed that the vacancy flips such that the density of states 
probed by the STM is the average of the rebonded and nonrebonded sides
of the vacancy (average of the two mirror configurations). The images
agree very well with the experiment. In particular, two major features ---
the nearly unchanged separation of the neighboring empty dangling bonds
[Fig. 3(c2)] and the depression of the neighboring
occupied dangling bonds [Fig. 3(c1)] --- are well reproduced. Close
inspection of Figs. 3(c) and 3(d) clearly demonstrates that the images of the
thermally flipping vacancy configuration explains the STM images best.
Note that the somewhat higher intensity in the experimental empty state images
is due to the local electrostatic potential arising from the positive chare
\cite{1} and that the small intensity of empty density of states above the
anions decays quickly into the vacuum such that the STM cannot probe it \cite{21}.

In conclusion, we have demonstrated a methodology to  determine the atomic 
structure and the
energy of  charge transfer levels in the band gap of surface defects,
by combining scanning tunneling microscopy,
photoelectron spectroscopy, and density-functional theory calculations.
From linking STM and PES data we find a charge transfer level 
(+/0) $0.75 \pm 0.1$ eV above the VBM for the example of P vacancies 
on InP(110) in good agreement with DFT calculations.
By comparing high resolution STM images with calculations, we
determine the vacancy to exhibit a rebonded nonsymmetric structure. 
The vacancy flips thermally between its two
mirror configurations. 
This localized vibrational mode leads to an apparently symmetric density
of
states in STM images, although the vacancy has a nonsymmetric atomic
structure. Similar measurements may help to understand on the atomic
scale the properties of a wide range of
materials, where defects play a key role. 

We thank Th.\ Chass\'e for assistance in an initial experiment, W. M\"onch
for valuable discussions, and K. H. Graf for technical support. This
work was supported by the Bundesministerium f\"ur Forschung und Technologie
under Grant No.\ 05SE8 OLA7 and by the Deutsche Forschungsgemeinschaft
through Sonderforschungsbereich 296 TP A4 and A5 and Grant No.\ Ne428/2-1.

%References


\begin{thebibliography}{20}
\bibitem{110} 
The theoretical results for the positively charged arsenic vacancy on
GaAs(110) exhibit a very close agreement to those for the respective P vacancy
on InP(110). 
Thus both vacancies can be expected to show a thermal flip motion resulting in
good agreement between STM simulation and experiment.
\end{thebibliography}
\end{document}